\documentstyle[aps,epsf]{revtex}
\newcommand{\be}{\begin{equation}}
\newcommand{\ee}{\end{equation}}
\newcommand{\Leq}[1]{\label{eq:#1}}
\newcommand{\Req}[1]{\ref{eq:#1}}
\newcommand{\bea}{\begin{eqnarray}}
\newcommand{\eea}{\end{eqnarray}}
\newcommand{\hh}{\tilde{h}}
\newcommand{\prt}{\partial}

\tighten
\begin{document}

\title{Quantum manifestation of L\'evy--type flights in a chaotic system }
\author{ A. Iomin$^{1,2}$ and G. M. Zaslavsky$^{2,3}$\\ 
$^1$ Department of Physics, Technion, Haifa 32000, Israel \\
$^2$Courant Institute of Mathematical Sciences, \\
New York University, 251 Mercer Str., NY 10012   \\
$^3$Department of Physics, New York University, \\
2-4 Washington Place, NY, NY 10003  }

\maketitle
\begin{abstract}
Semi--classical dynamics of quantum wave packets spreading is studied
for a kicked rotor. Quantum flights are established for a specific, 
``magic'' value of a chaos control parameter when the classical stickiness 
of trajectories is most effective. By studying of survival 
probability and distribution of the accelerations we identify the 
presence of quantum L\'evy--type flights.

PACS number(s): 05.45.Mt, 05.60.Gg
\end{abstract}

\vspace{3ex}

L\'{e}vy or L\'{e}vy-type processes occur in different processes related to 
chaotic dynamics (see reviews in \cite{1z,2z,3z,4z,5z}).
In Hamiltonian dynamics a phenomenon of stickiness of the chaotic 
trajectories
to island boundaries reveals an intermittent kinetics which can be partly
described by L\'{e}vy-type processes
and fractional kinetic equations \cite{6z,7z,zn}. One can find more 
applications in recent publications \cite{8z,9z}. Sticking islands can emerge 
from a tangent bifurcation \cite{mk83a,mk83b} or from periodic solutions of 
parabolic type \cite{10z,11z}. The phenomenon of stickiness to islands 
reveals in power 
dependence of the correlation functions, Poincar\'{e} recurrences 
distribution,
exit time distribution, etc. \cite{4z,mk83a,zen,12z,12zz}. Algebraic 
dependence of these distributions and moments is not universal; it 
depends on the models of intermittent dynamics, time intervals, domain of
parameters, etc. \cite{6z,12zz}, and even using a single characteristic 
exponent 
for the algebraic dependence is only a rough approximation to a much more
complicated multi-fractal structure of dynamics in chaotic sea in the 
presence of islands. 

Despite the complexity of chaotic dynamics with sticky islands, anomalous
diffusion in the form $ \langle |p|^{\alpha} \rangle \sim t^{\beta} $,
with momentum $p$ and appropriate exponents $\alpha$ and $\beta$, can
often be a good approximation, at least under specified conditions. The
case when $ \mu = 2\beta /\alpha > 1 $
is called super-diffusion, while the case $\alpha = 2$, $\beta = 1$
corresponds to a normal diffusion. There are different ways to explain
the emergence of super-diffusion. One of them, that works for the standard 
map or
the web map, is that a trajectory sticks to an accelerator mode island 
boundary
and travels for a while almost in the same way as a regular (non-chaotic)
trajectory of the accelerator mode \cite{zn,mk83a,zen,12zz}. As a result, 
one can observe long flights (L\'{e}vy-type flights)
in phase space, which corresponds to almost regular pieces of chaotic
trajectories. More precise definition of flights refers to the parts of
a trajectory with a Lyapunov exponent $\lambda$ fairly small comparing
to a typical value of $\lambda$ far from the island boundary layers.

In one way or another, the phenomenon of L\'{e}vy-type flights,
algebraic correlation decay, stickiness, etc., became objects of intensive
studies in quantum physics: L\'{e}vy-type processes
of scattering and random matrix distributions \cite{13z,kus} recoils 
distribution in the laser atoms cooling \cite{14z,15z}; anomalous transport 
in the quantum kicked rotor \cite{ott,bglr96,zs99,iz2} and in the Harper 
model 
\cite{iz1}; algebraic decay of the survival probability in different systems 
\cite{lbog92,cms99,cgm00} and in microwave ionization
of Rydberg atoms \cite{16z,bcms00}; fractal conductance fluctuations 
\cite{ketz96}.

In this paper we consider a quantum kicked rotor (QKR) system, in 
which the
influence of the sticking dynamics on quantum localization process was
already demonstrated in \cite{zs99} for a specific value of the chaos
control parameter $K=K^* = 6.908745\ldots$, called
``magic'' number.
It has been shown in \cite{zen} that the phenomenon of stickiness of a
trajectory to the hierarchical islands-around-islands structure is the
most efficient at this magic value of $ K $.
The trajectories from the chaotic sea can travel over long distances in 
the phase space
together with an island if the structure of the boundary islands chain 
(BIC) is sticky. The stickiness of a trajectory to the hierarchical 
islands-around-islands structure is intimately related to magnitude of the 
parameter $ K $. It has been shown in \cite{zen} that the effect of 
non-Gaussian chaotic dynamics (L\'evy--type process) becomes more 
observable when the BIC appears for special magic values of $ K $. 
In particular, for 
the standard map the value of $ K=K^*=6.908745 $ corresponds to the sequence
of islands $ 3-8-8-8-\dots $. This exact self-similarity of the islands
structure is also reflected in the scaling of the spatio-temporal 
characteristics of the every island chain generation.
A renormalization group was introduced in the form $ T'\rightarrow 
\lambda_TT,~~S'\rightarrow\lambda_SS ~~(\lambda_T>1,~\lambda_S<1) $, where
$ S,~S' $ are the areas and $ T,~T' $ are the periods of the last invariant 
curves inside of the islands, and these characteristics are related to the 
islands 
chain and its next generation. The scaling constants $ \lambda_T $ and
$ \lambda_S $ determine the transport exponent $ \mu=\frac{|\ln\lambda_S|}
{\ln\lambda_T} $, and this theoretical prediction is in a good agreement 
with numerics \cite{zen}. It should be noted that there are no strict
conditions on the hierarchy of islands in the case of anomalous 
transport, and
the sequences of values for the islands-around-islands periods can be
any numbers permitted by the possible resonances of the system. Each magic
value of $ K $ generates its own sequence of islands ({\em i.e.}, it is
$ 3-8-8-8-\dots $ for $ K=6.908745\dots $ \cite{zen}, or
$ 8-8-8-\dots $ for $ K=6.35933\dots $ and $ 5-11-11-11-\dots $
for $ K=6.476939\dots $ \cite{12zz}) and its own scaling constants 
$ \lambda_S $ and $ \lambda_T $. Therefore different hierarchical sequences
specify different deviations from the Gaussian transport, {\em i.e.}
normal diffusion, and different characteristic time scales when the 
anomalous properties of the diffusion can be observed in both classical 
and quantum cases.

An essential increase in the level occupation amplitudes
$|f_n |$ of the order of $10^5$ has been found for the magic value
$K^*$ for strong quantum \cite{zs99} and semi-classical \cite{iz2} cases.
Anomalous diffusion \cite{bglr96} and the quantum tunneling from the
acceleration mode islands \cite{ott} for QKR have been studied as well.
These effects are implicit indications of L\'{e}vy-type flights
taking place in the quantum system due to the classical sticking dynamics.
Accelerator mode islands move a quantum wave packet in the
energy levels space of the unperturbed system  faster
than in the case of global chaos, since the Lyapunov exponents
in the vicinity of these islands are considerably smaller than in the chaotic
sea. Hence a trajectory sticks to this hierarchical area for long time.
The question of direct observation of these flights is still open, and
we consider this problem here.

We study numerically the anomalous transport for the Hamiltonian of the QKR
\be
 H=\frac{p^2}{2}+K\cos q\delta_T(t) ,
\label{eq:qkr}
\ee
where  $ p=-i\hh\frac{\prt}{\prt q} $ is the angular momentum and
$ q\in[0,2\pi] $ is the angle, while $ \hh $ is the dimensionless Planck 
constant,
$ K $ is the control parameter, and $ \delta_T(t)=\sum_n\delta(t-n) $ is a 
periodic train of $ \delta $ -- kicks with the period $ T=1 $.
The quantum map
\begin{equation}
\Psi(t+T)=\hat{U}(T)\Psi(t)
\label{eq:for2}
\end{equation}
is produced by the following evolution operator over the period $ T $ :
\be
 \hat{U}=\exp\{-i\hh \hat{n}^2/2\}\exp\{-i(K/\hh)\cos q\} .
\label{eq:hatU}
\ee
 Below we study the evolution of the complex amplitudes for the  
expansion of the wave function $ \Psi $ in terms of the 
unperturbed system basis $ e^{inq} $,
\be
\Psi(t)=\Psi(q,t)=\sum_{n=-\infty}^{\infty}f_n(t)e^{inq} 
\Leq{Psit}
\ee
Here $ |f_n(t)|^2 $ 
is the probability of level occupation at time $ t $ which determines the 
spreading of the energy growth over the unperturbed spectrum. It enables 
us to study the  
evolution of survival or staying probability in an energy region with 
momentum range $ \Delta p= \hh\Delta N $, which is not necessarily 
compact, and determined  without constraints on the boundaries:
\begin{equation}
P(t)=\sum_{n\in\Delta N}|f_n(t)|^2.
\label{eq:for4}
\end{equation}
For iteration of the map (\ref{eq:for2}) we use the standard technique of the 
fast Fourier transform with up to $ 2^{18} $ angular momentum eigenstates.
We study also the time evolution of the second moment $ \sigma_2=
\langle p^2\rangle =\hh^2\sum_n|f_n|^2n^2 $ as a function of $ K $ and 
the initial conditions.
The initial conditions for the amplitudes of level occupation are $ f_n(t=0)=
\delta_{n,n'} $ where $ n'=0 $ corresponds to the initial population inside 
acceleration islands, while $ n'=\pm 31 $ correspond to the initial population
in chaotic sea with $ p\approx\pm\frac{3\pi}{8} $ for $ \hh=0.0396 $. The 
values of $ \hh $ are irrational numbers or rational approximations of 
irrational numbers such that quantum resonances with $ \sigma_2\sim t^2 $ 
are avoided. 

We started the study of the wave packet spreading for $ 
f_n(t=0)=\delta_{n,0} $.
When the chaos parameter $ K>1 $ approaches the magic number 
$ K^*=6.908745\dots $ the anomalous super-diffusion takes place in the 
energy space. In this case, sticking dynamics inside of the hierarchical 
islands around islands  structure 
spreads the initial wave packet very fast. This leads to the anomalously 
diffusive growth of the second moment $ \sigma_2(t)\sim t^{\beta} $, where 
$\beta\approx 1.57\pm 0.27 $ in the time interval $ t\in [0,300]  $. Then 
the transport exponent  changes to diffusive one $\beta\approx 0.98\pm 
0.04 $ 
determined in the time interval $ t\in [400, 1000] $ (see Fig.\ 1).
The diffusion coefficient $ D \sim 10^3 $ is anomalously large and 
differs essentially from normal diffusion  coefficient $ K^2/4\ll 
D $.
 
The super--diffusion for $ K=K^* $ is also determined by an increase 
of $ |f_n| $ by more than $10^{10} $ times in comparison with
$ |f_n| $ corresponding to usual diffusion with $ K=9.69 $
(see Fig.\ 2). This differences can be explained by quantum flights, 
which reflect L\'evy--type flights in the classical limit. To illuminate these 
flights we compute survival probability in periodic energy level regions 
taken in periodic classical cells of the size $ 2\pi\times 2\pi $. To that 
end 
$ \frac{\hh}{2\pi}=\frac{1}{157+gm} $ is determined by a rational 
approximation of the 
golden mean value: $ gm=55/34 $, and one obtains for the dimensionless Planck 
constant $\hh=2\pi\cdot R/Q=2\pi\cdot 34/5393 $.
Therefore $\hh $ is commensurate with energy level spacing with the 
period of 5393 levels, which corresponds to the 34 cells of the size 
$ 2\pi $ on a  cylinder.
 One 
takes 40 levels in  every periodic cell and studies the change of $ P(t) $ 
in (\ref{eq:for4}) with time, where 
\be
\Delta N=\bigcup_{l=-a}^a[l\cdot Q+sgn(l),l\cdot Q+sgn(l)\cdot 40] 
\label{eq:deltaN}
\ee
and $ a $ is an integer part of the 
ratio $ N/2Q $. The result of the numerical calculation is shown in Fig.\ 3.
Every passage of the wave packet through the periodic cell contributes to
the increase of survival probability  $  P(t) $. 
This causes the spikes of $  P(t) $.
The period between any two spikes is equal to the number of periodic cells: 
$ \Delta t =R=34 $. This means that the part of the wave packet which  
sticks to the BIC 
passes through one cell per one iteration, {\em i.e.} it moves 
ballistically. 
Amplitudes of the spikes decay exponentially with time by the following 
expression $ \sim \exp(-0.005t) $. 
It relates to the quantum tunneling through cantori \cite{ott,grr96,note1}.
This behavior of quantum tunneling from the accelerator mode to 
the chaotic sea is in a good agreement with the result obtained in 
\cite{ott,note2}. 

The spikes disappear if  the initial conditions are chosen inside of the 
chaotic sea with $ n'=\pm 31 $ for $ f_n(t=0)=\delta_{n.n'} $.
Nevertheless, super-diffusion due to the sticking to BIC dynamics takes 
place  for $ K=K^* $ with  $ \sigma_2 \sim t^{\beta} $, where 
$ \beta\approx 1.39\pm 0.16  $ .
Also, in comparison with $ K=9.69 $, an increase of order of magnitude in 
the level occupation amplitudes $ |f_n| $ is obtained.
It is the same order as shown in Fig.\ 2. 
These differences unambiguously indicate the presence of flights. 
To pinpoint these quantum flights, we calculate a set of the
survival probabilities $ P_k(t) $ constructed over a set of inclosed 
compact  regions of the energy levels $ \Delta N(k)=N_0k $, where $ N_0=3^6 $ 
and $ k=1\div 10 $ are integer numbers. 
Let us denote the time intervals  which are necessary for the
wave packet to reach the boundaries of $ \Delta N(k) $ as $ t_k $. In our
numerical calculations $ t_k $ is the number of iterations.
We found a linear growth of the times $ t_k $ which is proportional to 
the increase of size of the regions $ \Delta N(k) $, such that 
$ t_k\sim \Delta N(k) $ (see Fig.\ 4(c)).
It means that the part of the wave packet which reaches the boundary
of $ \Delta N(k) $ at the moment $ t_k $ moves like in an acceleration mode: 
$ \Delta p\sim \hh\Delta N\sim \Delta t $. As far as the diffusive 
spreading of the wave packet for $ K=9.69 $ goes, one can see 
in Fig.\ 4(a) that in this case $ t_k $ is not proportional to
$ \Delta N(k) $. Another important
characteristic of the wave packet evolution is the decay rate of the 
survival probability.
Starting  from $ t_k $ the survival probability $ P_k(t\geq t_k) $ in 
(\ref{eq:for4}) decays in correspondence with the rate of the 
spreading of the wave packet.
The decay rate depends on $ K $. In the chaotic case when $ 
K=9.69 $ only the  tail of the wave packet moves ballistically. It 
corresponds
to the transitions between any two levels $ n_1 $ and $ n_2 $ such that 
$ |n_1-n_2|\sim K/\hh $. In this case the matrix elements of the 
evolution operator (\ref{eq:hatU}), $ U_{n_1,n_2}=e^{-in_1^2\hh/2} 
J_{n_1-n_2}(K/\hh)\sim \sqrt{\hh/K} $, where $ J_m(z) $ is the Bessel 
function of the first kind, are small and uncorrelated random quantities. 
Their total contribution to the transition
is summing up to an exponentially small value, which is reflected by the 
exponentially small values of $ |f_n| $ in Fig.\ 2(a). 
Therefore, $ P_k(t) $  decay gradually and slowly with time as it is shown 
in 4(a,c). For $ K=K^* $
the values of $ P_k(t) $ fall down abruptly {\it e.i.} much faster then in 
the chaotic case, which is illustrated in Fig.\ 4(b,d). It means that 
there is a part of the wave packet that moves ballistically together with 
the tail. Obviously that this part determines the decay of $ P_k(t) $. 
Therefore, in case of mixed phase space the hierarchical 
sticking islands structure leads to ballistic flights. Fast change of $ 
P_k(t)
 $ in Figs.\ 4(b,d) relates to these quantum flights, while
the slow changes of $ P_k(t) $ in Figs.\ 4(a,c) relate to normal 
diffusion.

It should be noted that the classical nature of the phase space
reflects also in pure quantum characteristics of the evolution 
operator  $ \hat{U} $ (\Req{hatU}). Let us show that the matrix elements 
corresponding to flights, such that $ |n_1-n_2|\sim K/\hh $ are much 
stronger correlated for $ K\approx 2\pi $ than for $ K=9.69 $.
We consider the following correlation function:
\be
\rho(\eta)=\sum_n\langle\langle U_{m,n}U_{n,m+\eta}^*{\rangle\rangle}_m,
\Leq{corfun1}
\ee
where $ \langle\langle\dots{\rangle\rangle}_m\equiv\frac{1}{L}\sum_{m=1}^L
\dots $ means averaging over $ m $ in some energy levels region, such that
$ 1\ll L\ll 1/\hh $. By using the explicit form of the matrix elements 
$ U_{m,n} $, one obtains for the correlation function
\be
\rho(\eta)=\frac{1}{L}e^{-i\hh\eta^2/2}J_{\eta}(2K/\hh)
\frac{e^{i\hh\eta L}-1}{1-e^{i\hh\eta}}.
\Leq{corfun2}
\ee
Let us take $ \eta=[k/\hh]=K/\hh-\{K/\hh\} $, where
$ [Z] $ is the integer part, while $ \{Z\} $ is the fractional
part of $ Z $. Therefore $ \hh\eta=K-\hh\{K/\hh\}=K-\varepsilon $,
and the condition $ L\varepsilon\ll 1 $ is also fulfilled. Hence,
it is easy to see from (\Req{corfun2}), that if $ K=2\pi+\varepsilon $
(or $ K=2\pi n+\varepsilon $), then $ \rho(\eta)\equiv
\rho_{acc}([K/\hh])\sim J_{[K/\hh]}(2[K/\hh]) $ is 
the correlation function with presence the acceleration mode motion
in the classical counterpart. It is also easy to see for the 
global chaos case (including $ K=9.69 $), that the correlation function
satisfies $ \rho(\eta)\equiv\rho_{diff}([K/\hh]=\frac{1}{L}
\rho_{acc}([K/\hh])\ll\rho_{acc}([K/\hh]) $.
It means that for the diffusive case without any stability islands the 
matrix elements  are fairly less correlated than in the case of anomalous 
diffusion, in which the acceleration mode islands (with $ K=2\pi n 
+\varepsilon $) play crucial role for both classical and quantum transport 
and quantum localization of anomalous diffusion \cite{iz2}.
It should be noted that for $ K=2\pi $ the motion is diffusive both for
classical and quantum case. Only for $ 2\pi<K<7 $ the anomalous 
diffusion has been found. Similarly to our result,
the anomalous diffusion for QKR has been found 
in \cite{ott} and \cite{bglr96} for $ K=2\pi\cdot 1.03 $ and 
$ K=6.9115 $ correspondingly. This range of $ K $ relates to bifurcate 
emerging and existing of the acceleration mode islands. That fact that
the quantum correlation functions (\Req{corfun1}) reflects the sticking
dynamics in the vicinity of the hierarchical structure of these islands
relates to the fundamental correspondence principle by which
the semi--classical wave packet evolves alone an individual trajectory
during the quantum correspondence time $ \tau_{h} $ \cite{bz78,zs99,iz2}.
The $ \hh $-scaling for this time is algebraic, $ \tau_{h}\sim
\hh^{\frac{-1}{1+\beta}} $ \cite{iz3} for the acceleration mode dynamics.
This result corresponds to general consideration of quantum dynamics 
with mixed phase space \cite{cms99,ketzm}

To complete this analysis of quantum flights, we consider the 
acceleration statistics. These values are defined as the mean square 
momentum changes at every iteration,
\be
 \delta p(t)\equiv\langle (p_{t+1}-p_t)^2\rangle/K^2=\langle\sin^2 
q_t\rangle,
\Leq{dpt1}
\ee
where $ \langle\dots\rangle $ means the quantum average. It can 
be expressed by the level occupation  amplitudes in the following form:
\be
 \delta p(t)=
\sum_n\Big[\frac{1}{2}|f_n(t)|^2-\frac{1}{4}f_n^*f_{n-2}-\frac{1}{4}
f_n^*f_{n+2}\Big].
\Leq{dpt2}
\ee
One should note that the dynamical consideration is restricted in time 
because of the fast spreading of the wave packet. This leads to the fast 
overfill  of energy levels region  with the fixed size of $ N=2^{18} $ 
levels. Therefore, the maximal number of iterations is restricted and equals 
to 1000 for $ \hh=0.0399 $. Dynamical calculations of $ f_n(t) $ are carried 
out for a set of initial conditions, arbitrary chosen inside the chaotic sea
to obtain an appropriate statistics.

Let us consider probability distribution function  $ W_K(\delta p) $ 
of accelerations $ \delta p $. For the white noise $ W_K(\delta p) $ is 
Gaussian, while for a process which corresponds to the fractional kinetics, 
one expects $ W_K(\delta p)\sim1/(\delta p)^l $, where 
$|\delta p|\rightarrow\infty $.
The exponent $ l=\log_{10}(W_K(\delta p)/\log_{10}(\delta p) $ determines to 
what kind of statistics  $ W_K(\delta p) $ belongs.
When $ l $ is small enough, 
we associate the distribution $ W_K(\delta p) $ with the L\'evy--type 
probability distribution function,
and vice versa, if $ l $ is large, $ W_K(\delta p) $ is associated with
the Gaussian one, since one cannot distinguish between large $ l $ and 
exponentially small values of $ W_K(\delta p) $. The distribution of 
accelerations 
$ \Delta p=\delta p-\overline{\delta p} $ for the set of 36 different initial 
conditions
with $ 31<|n_0| <70 $, arbitrary chosen inside the chaotic region is 
plotted in 
Fig.\ 5. Here $ \overline{\delta p} $ is the mean value averaged over $ t $.

The probability distribution functions are asymmetric and we compare only 
the right tails of the chaotic dynamics with $ K=9.69 $ and  the sticky case 
with $ K=K^* $. The distribution function for chaotic, non--sticky case 
has a short right tail with a large slope of $ l\approx 9.5 $.
This magnitude  of $ l $ is large enough, and one concludes that the 
semi-classical diffusion gives rise to Gaussian distribution for the 
accelerations. The distribution function for $ K=K^* $ differs essentially 
from the Gaussian distribution. As it is shown in Fig.\ 5, it is narrower 
with a longer tail and the slope is of the order of $ l\approx 4.3 $.
These differences enable us to conclude that $ W_{K=K^*}(\Delta p)  $ is 
the L\'evy-- type distribution and relates to the quantum flights 
which are due to the effect of stickiness of the semi--classical wave 
packet to the classical hierarchical islands--around--islands structure.

This research is supported by U.S. Department of Navy Grants 
No N00014-96-10055, No N00014-97-1-0426 and by U.S. Department of Energy
Grant No DE-FG02-92-ER54184. A.I. was also supported by the Israel
Science Foundation.

\newpage
\section*{Figure Captions}
\begin{description}
\item[Fig.~1]
Time evolution of the second moment $\sigma_2 $ for initial population 
inside the acceleration mode island: $ n'=0 $, and  $\hh=\frac{2\pi\cdot 
34}{5393} $, where (a) $ K=9.69 $, and (b) $ K=K^* $.
\item[Fig.~2]
Level occupation amplitudes $ |f_n| $ vs $ n $ for $ K=9.69 $ (a) and
$ K=K^* $ (b) after 1000 iterations. The slop of the linear plot for 
the energy region $ (6\div 12)\times 10^4 $ is $ 1.4\cdot 10^{-5} $. 

\item[Fig.~3]
Semi-log  plot of the survival probability $ P(t) $ vs $ t $ for
periodic cells with the period $ 64\pi $. The plots are: (a) for 
$ K=9.69 $, and (b) for $ K=K^* $. 

\item[Fig.~4]
A set of survival probabilities $ P_k(t) $ vs $ t $ for $ K=9.69 $
(a,c) and $ K=K^* $ (b,d) for variety of energy regions 
$ \Delta N (k)=3^6\cdot k $ and $ k=1\div 10 $ and $ \hh=2\pi/(157+gm) $  
Plots (c) and (d)  zoom in of (a) and (b) respectively for 
$ k=3 $.

\item[Fig.~5]
 (upper plots) Probability distribution functions $ W_K(\Delta p) $ 
for 120 boxes and the centered by $ \Delta p =\delta p-\overline{\delta p} $, 
where $ \overline{\delta p} $ is the mean value with $ K=9.69 $ for (a);
and $ K=K^* $ for (b). 
(bottom plots) Slops for the right tails of the probability distribution 
functions, where (a) corresponds
to $ K=9.69 $ and the slop is $ 9.51\pm 0.02 $ for $ \Delta p \in
[0.0111,0.0146] $; (b) corresponds to $ K=K^* $
and the slop is  $ 4.31\pm 0.06 $ for $ \Delta p \in [0.0109,0.0156] $. 

\end{description}


\begin{references}
\bibitem{1z} E. W. Montroll and M. F. Shlesinger, in
``Studies in Statistical Mechanics'', edited by J. Leibowitz and
E. Montroll (North-Holland, Amsterdam, 1984), vol. 11, p. 1.

\bibitem{2z} J. P. Bouchand and A. Georges, Phys. Rep. {\bf 195}, 127 (1990).

\bibitem{3z} ``L\'{e}vy Flights and Related Topics in Physics'', edited by 
M. Shlesinger, G. M. Zaslavsky, and U. Frisch (Springer-Verlag, Heidelberg, 
1995).

\bibitem{4z} M. F. Shlesinger, G. M. Zaslavsky, and J. Klafter,
Nature {\bf 363}, 31 (1993).

\bibitem{5z} T. Geisel, in \cite{3z}, p. 153.

\bibitem{6z} V. Afanasiev, R. Z. Sagdeev, and G. M. Zaslavsky,
Chaos {\bf 1}, 143 (1991).

\bibitem{7z} G. M. Zaslavsky, Physica D {\bf 76}, 110 (1994); Chaos 
{\bf 4}, 25 (1994).

\bibitem{zn} G. M. Zaslavsky and B. A. Niyazov, Phys. Rep. {\bf 283},
73 (1997).

\bibitem{8z} P. Leboeuf, Physica D {\bf 116}, 8 (1998);
S. S. Abdullaev and K. H. Spatschek, Phys. Rev. E {\bf 60}, R6287(1999);

\bibitem{9z} G. Zimbardo, P. Veltri, and P. Pommois,
Phys. Rev. E {\bf 61}, 1940 (2000).

\bibitem{mk83a} J. D. Meiss, {\em et al}., Physica D {\bf 6}, 375 (1983).

\bibitem{mk83b} J. D. Meiss, Phys. Rev. A {\bf 34}, 2375 (1986);
Rev. Mod. Phys. {\bf 64}, 795 (1992).

\bibitem{10z} V. Melnikov, in ``Chaos, Transport, and Plasma Physics'', 
edited by S. Benkadda, F. Doveil, and Y. Elskens (World Scientific, 
Singapore, 1994), p. 126. 

\bibitem{11z} V. Rom-Kedar and G. M. Zaslavsky,
Chaos {\bf 9}, 6973 (1999).

\bibitem{zen}  G. M. Zaslavsky, M. Edelman, and B. A. Niyazov, Chaos 
{\bf 7}, 159 (1997).

\bibitem{12z} B. V. Chirikov and D. L. Shepeliansky,Physica D {\bf 13}, 395 
(1984); B. V. Chirikov, Chaos, Solitons, Fractals {\bf 1}, 79 (1991).

\bibitem{12zz} G. M. Zaslavsky and M. Edelman, Chaos {\bf 10}, 135 (2000);
S. Benkadda, {\it et al.}, Phys. Rev. E {\bf 55}, 4909 (1997);
{\bf 59}, 3761 (1999).

\bibitem{13z} D. Kusnezov, Phys. Rev. Lett. {\bf 72}, 1990 (1994);
D. Kusnezov and C. H. Lewenkopf, Phys. Rev. E {\bf 53}, 2283 (1999).

\bibitem{kus}  D. Kusnezov, A. Bulgak, and G. D. Dang,  Phys. Rev. Lett.
{\bf 82}, 1136 (1999).

\bibitem{14z} F. Bardou, J. P. Bouchaud, O. Emile, A. Aspect, and C. 
Cohen-Tannoudji, Phys. Rev. Lett. {\bf 72}, 203 (1994);
B. Saubam\'{e}a, M. Leduc, and C. Cohen-Tannoudji,
Phys. Rev. Lett. {\bf 83}, 3796 (1999).

\bibitem{15z} S. Schaufler, W. P. Schleich, and V. P. Yakovlev,
Phys. Rev. Lett. {\bf 83}, 3162 (1999).

\bibitem{ott} J. D. Hanson, E. Ott, and T. M. Antonsen,Jr., 
Phys. Rev. A {\bf 29}, 819 (1984).

\bibitem{bglr96} R. Roncaglia, L. Bonci, B. J. West, and P. Grigolini,
Phys. Rev. E {\bf 51}, 5524 (1995); L. Bonci, P. Grigolini, A. Laux, and 
R. Roncaglia, Phys. Rev. A {\bf 54}, 112 (1996).

\bibitem{zs99} B. Sundaram and G. M. Zaslavsky, Phys. Rev. E {\bf 59}, 7231
(1999).

\bibitem{iz2} A. Iomin and G. M. Zaslavsky, Chaos {\bf 10}, 147 (2000).

\bibitem{iz1} A. Iomin and G. M. Zaslavsky, Phys. Rev. E {\bf 60}, 7580
(1999).

\bibitem{lbog92} Y.-C. Lai, R. Bl\"umel, E. Ott, and C. Grebogi, Phys. Rev.
Let. {\bf 68}, 3491 (1992).

\bibitem{cms99} G. Casati, G. Maspero, and D. L. Shepelyansky, Phys. Rev.
Let. {\bf 82}, 524 (2000). 

\bibitem{cgm00} G. Casati, I. Guarneri, and  G. Maspero, Phys. Rev.
Let. {\bf 84}, 63 (2000).

\bibitem{16z} A. Buchleiner, {\it et al.},
Phys. Rev. Lett. {\bf 75}, 3818 (1995).

\bibitem{bcms00} G. Beneti, G. Casati, G. Maspero, and D. L. Shepelyansky,
cond-mat/9911200.

\bibitem{ketz96} R. Ketzmerick, Phys. Rev B {\bf 54}, 10841 (1996).

\bibitem{grr96} T. Geisel, G. Radons, and J. Rubner, Phys. Rev. Let. 
{\bf 57}, 2883 (1986).

\bibitem{note1} Our result is in a good agreement with \cite{ott},
where the quantum tunneling from the acceleration mode island has been 
studied for variety of $ \hh $ and $ K=2\pi(1.03)\approx 6.4717 $.
Using \cite{ott}  one obtains for $ \hh=\frac{2\pi\cdot 34}{5393} $
that the decay rate due to quantum tunneling is $ ~\exp(-5.3)\approx 0.005 $.

\bibitem{note2} We thank E. Ott for drawing our attention to this result.

\bibitem{bz78} G. P. Berman and G. M. Zaslavsky, Physica A {\bf 91}, 450
(1978).

\bibitem{iz3} A. Iomin and G. M. Zaslavsky, unpublished.

\bibitem{ketzm} R. Ketzmerick, L. Hufnagel, F. Steinbach, and M. Weiss,
Phys. Rev. Lett. {\bf 85}, 1214 (2000); L. Hufnagel, R. Ketzmerick, and
M. Weiss, cond-mat/0009010.
\end{references}
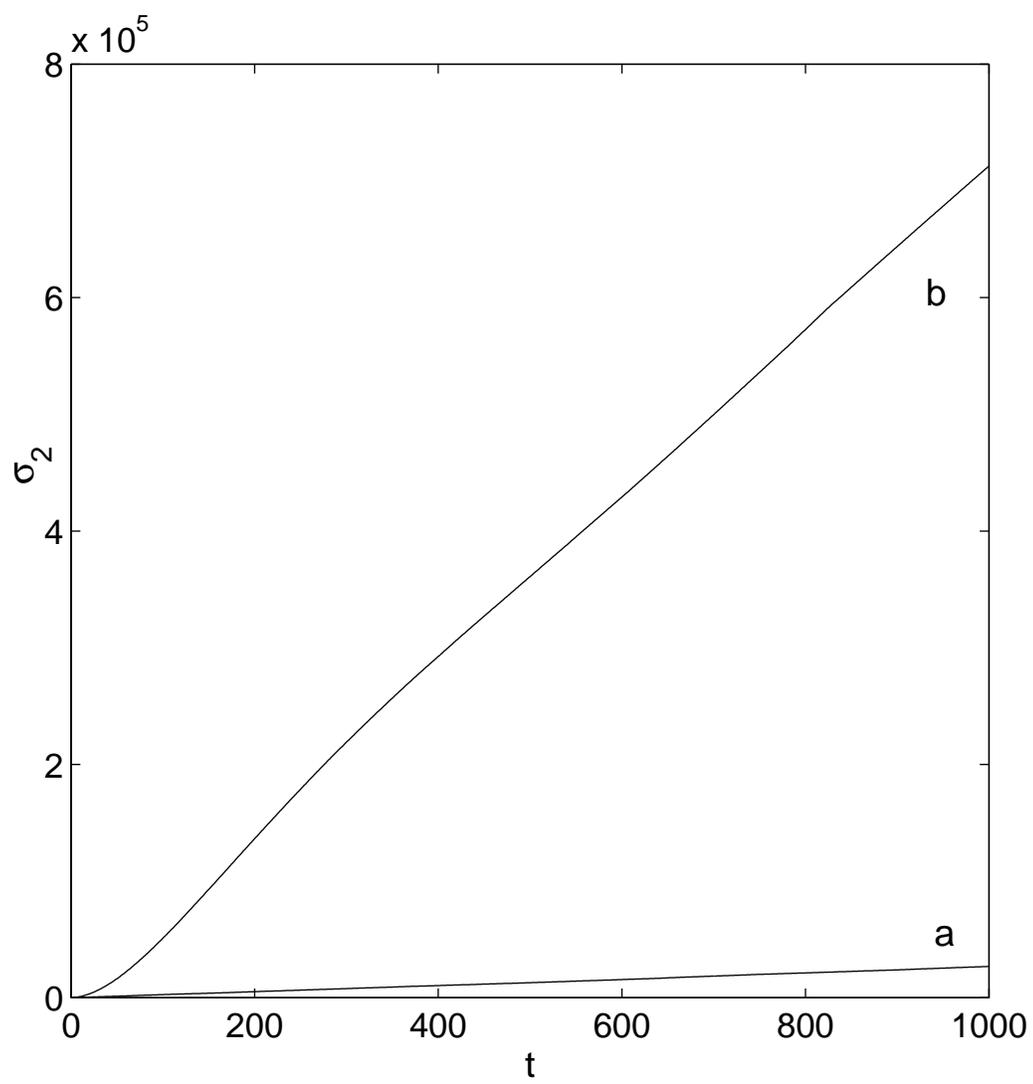
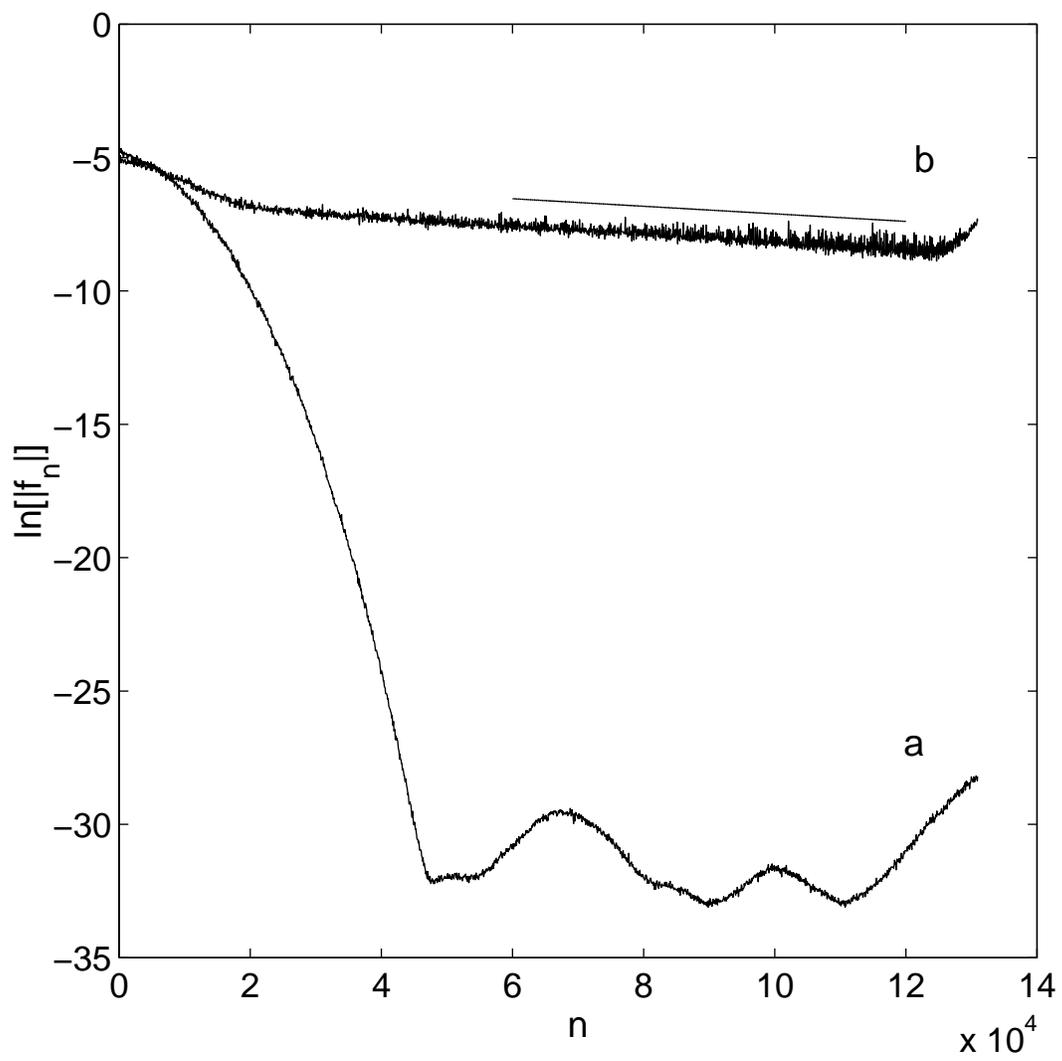
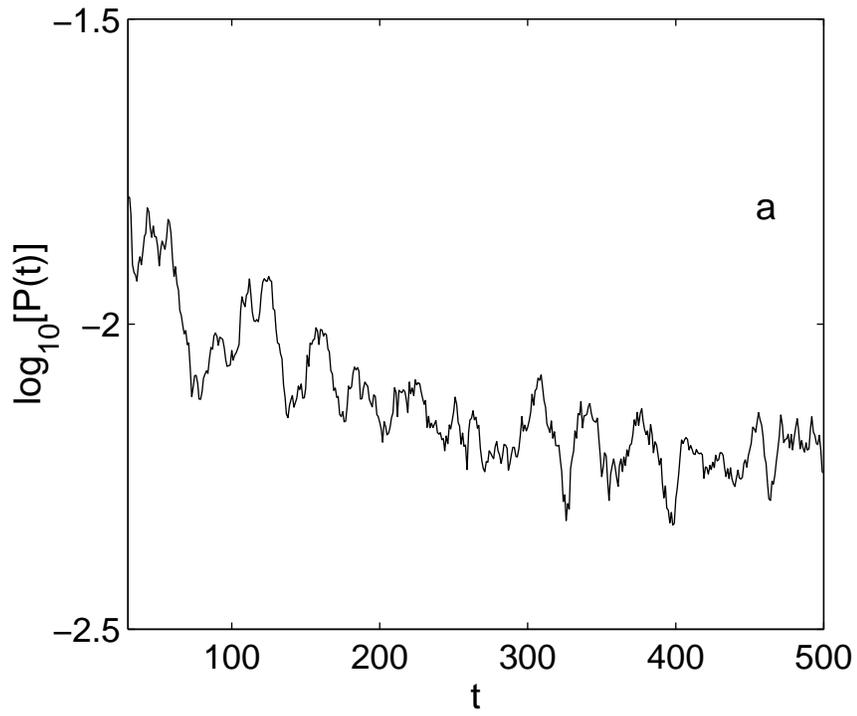
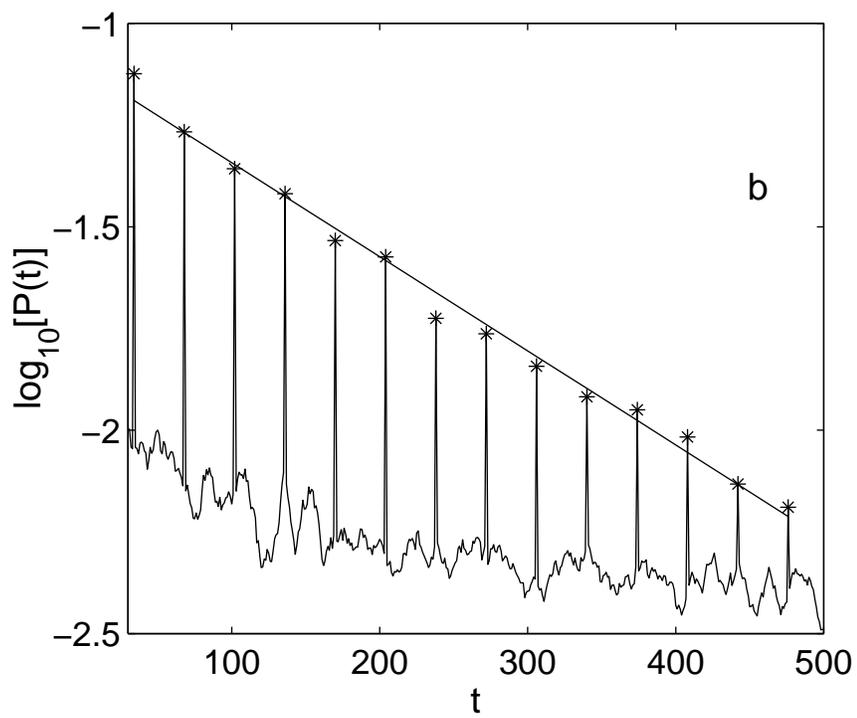
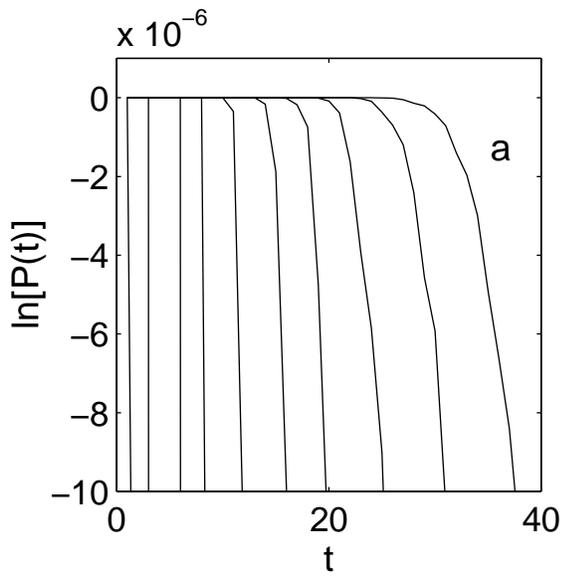
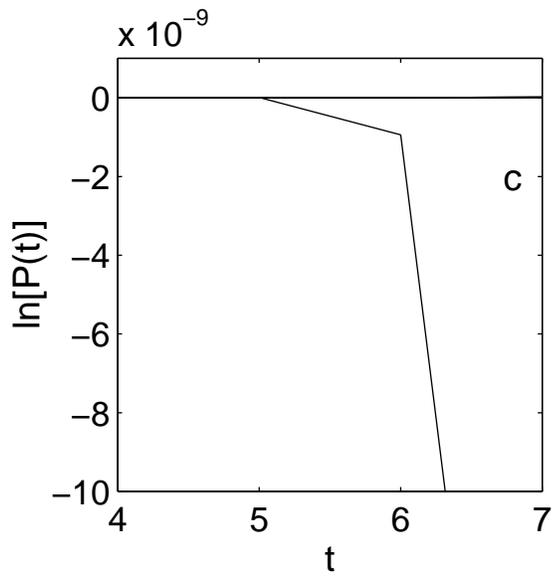
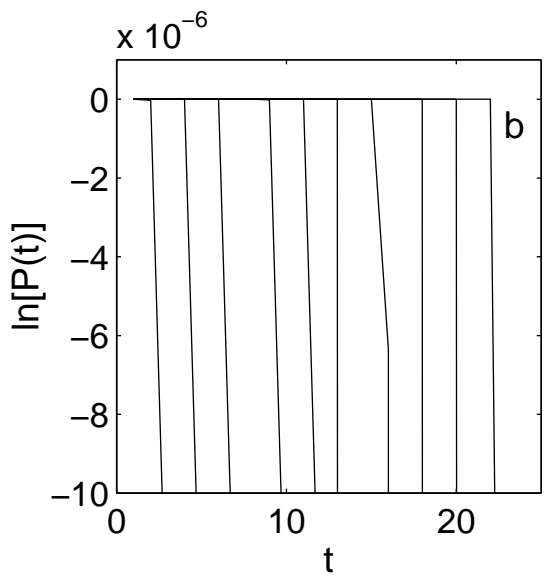
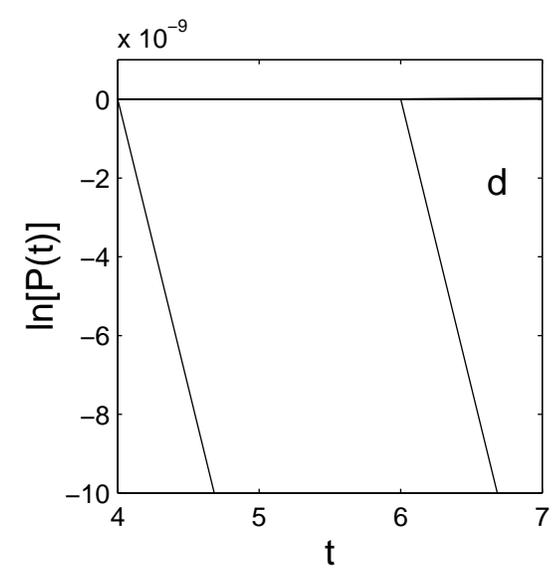
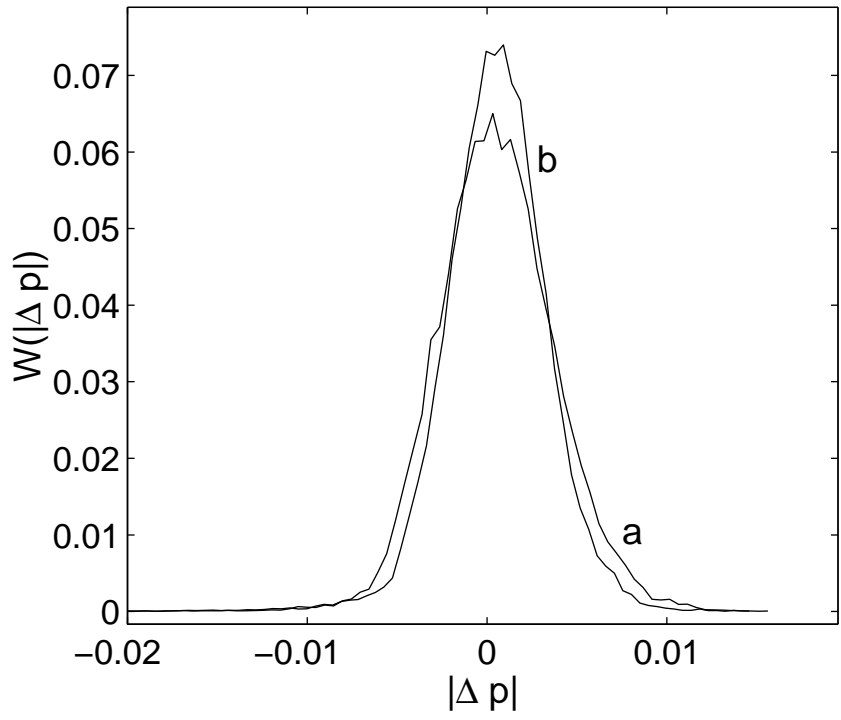
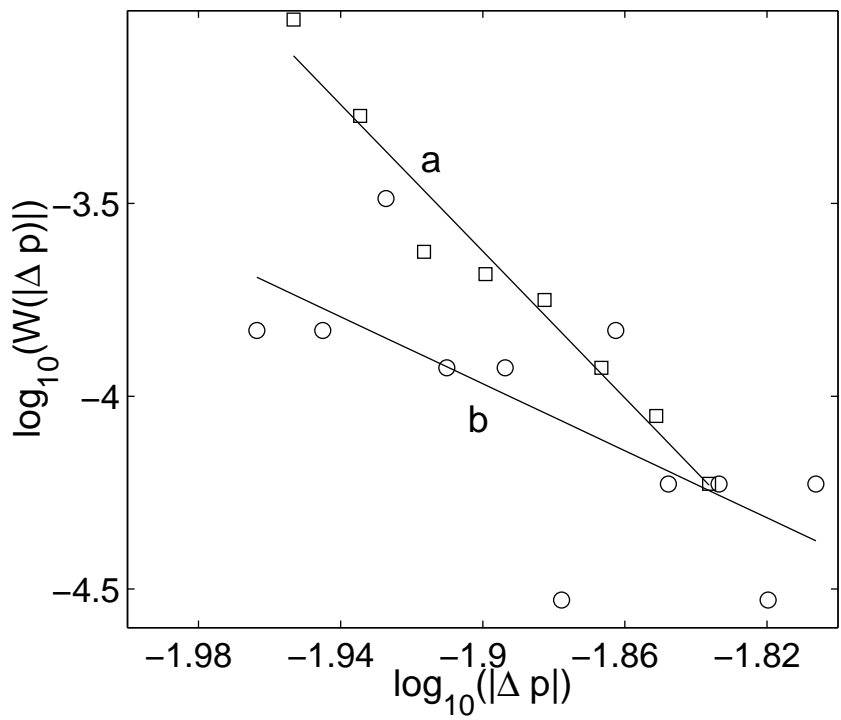
\end{document}